\documentclass[notoc]{JHEP} % pas draft
\usepackage{amsmath}
\usepackage{epsfig}
\usepackage{amssymb,amsfonts}
\newcommand{\BZ}{\mathbb{Z}}

\def\sqr#1#2{{
\vcenter{\vbox{\hrule height.#2pt
\hbox{\vrule width.#2pt height#1pt \kern#1pt
\vrule width.#2pt}
\hrule height.#2pt}}}}

\title{Flux Stabilization of D-branes}

\author{C. Bachas\\
 Laboratoire de Physique Th{\'e}orique
de l' Ecole Normale Sup{\'e}rieure\thanks{Unit{\'e} mixte  du
CNRS et de l ENS,  
UMR 8549.} \\ 
24  rue Lhomond, F-75231 Paris Cedex 05, France\\
\email{bachas@physique.ens.fr}}

\author{M. Douglas\\
 Department of Physics and Astronomy, Rutgers University\\
Piscataway, New Jersey 08855-0849, USA\\
and \\
I.H.E.S. \\
Le Bois-Marie, F-91440 Burs-sur-Yvette, France \\
\email{mrd@strings.rutgers.edu} }

\author{C. Schweigert\\
LPTHE, Universit{\'e} Pierre et Marie Curie, PARIS VI\\
Tour 16, 4 Place Jussieu, F-75252 Paris Cedex 05,  France\\
\email{schweige@lpthe.jussieu.fr}}

\abstract{We explain how  D-branes on group manifolds are stabilized against 
shrinking by quantized worldvolume $U(1)$ fluxes. Starting from the
Born-Infeld action in the case of the $SU(2)$ group manifold we derive
the masses, multiplicities  and spectrum of small fluctuations of these branes,
and show that they agree exactly with the predictions of Conformal Field 
Theory, to all orders in the $\alpha^\prime$ expansion. We discuss the 
generalization to other groups and comment on an apparent paradox: why are 
the `RR charges' of these branes not quantized?}

\preprint{LPTENS-00/08\\
PAR-LPTHE 00-10 \\
hep-th/0003037}

\begin{document}
\section{Introduction}

String compactifications on group manifolds have long been of interest.
Their world-sheet theories are the exactly solvable
Wess-Zumino-Witten (WZW) conformal field theories and thus stringy
effects can be understood in detail.
The case of primary interest in superstring theory
is the $SU(2)$ group manifold,
because the near-horizon geometry of $k$ coincident
NS fivebranes is a direct product
including $S^3$, and the corresponding CFT is a product of
supersymmetric $SU(2)$ level $k$, a ``Feigin-Fuchs superfield'',  
and six free superfields \cite{CHS}. Another exact supersymmetric string
background \cite{ABS}  is $S^3\times AdS_3$ -- this  corresponds  to the  CFT 
of two  supersymmetric WZW models,  one for the $SU(2)$ and one for 
the $SL(2,R)$ group manifold, and it   describes the near-horizon
geometry of intersecting branes \cite{CT,BPS}.

D-branes in group manifolds  have been studied in a series of papers
\cite{PSS,KS,KO,BS,AS,FS,ARS,BPPZ,S,GP}\  and the basic story,  at least
for the compact case, is fairly well understood. The natural
boundary conditions (those for which the gluing can be expressed in terms
of an automorphism $\omega$ of the current algebra) can be classified purely
in CFT terms. In the case of trivial gluing, $\omega=1$, Cardy's general
theory \cite{C}, puts them in one-to-one correspondence with primary fields.
\footnote{Cardy's theory can be generalized to non-trivial $\omega$;
in this case, one obtains a correspondence of boundary conditions with
primary fields in twisted sectors of appropriate orbifold theories,
\cite{FS}.}
The results turn out to be geometrical~: \cite{AS,FS}
an allowed boundary condition corresponds to a D-brane wrapped on
an allowed (twisted) conjugacy class of the group.
The only sign of the underlying CFT is a  quantization condition
on the allowed (twisted) conjugacy classes.
For example, in the $SU(2)$ level $k$
model D-branes can wrap on $k-1$ distinct $S^2$'s around any point
(these are subject to a $\BZ_2$ identification).
There is also a D$0$-brane, to complete the spectrum (there is no D$3$
because $H\ne 0$ \cite{witten}).

One then can study the world-volume theories of these branes by
classifying massless modes and computing interactions.  The results
show an intriguing parallel with the noncommutative torus in
that the algebra of open string primary fields (for large $k$ but
finite conjugacy class) is the algebra of the ``fuzzy sphere,''
the natural quantization of the sphere \cite{ARS}.

However, there is a more elementary question one might ask first.
Why is it that branes wrapped about spheres which are not minimal
volume surfaces are stable at all ?

It is easiest to check that the other boundary conditions
lead to stable branes by considering
the parallel with the boundary state describing a D$2$-brane ending
on an NS fivebrane.  This boundary state is a tensor product of ``D$0$''
in the WZW sector with ``Neumann'' in the linear dilaton sector and in
two of the Minkowski dimensions.  It preserves half of the supersymmetry
of the fivebrane theory.
If $k>1$, the WZW component of the boundary state can be replaced with
a different WZW boundary condition, leaving everything else unchanged.
In particular the supersymmetry of the object is unchanged, so it is
stable.  The geometrical interpretation of these objects is ``conical''
D$4$-branes again wrapping a non-minimal $S^2$.  Thus our question is
appropriate.

It seems clear that this stability is linked with the origin of the
quantization condition, and there are two ways one might try to explain
this.  One might imagine that the integrated 
NS two-form potential $\int {\hat B}$ takes
quantized values on a D-brane world-sheet. This would 
imply  an implausible non-local constraint on the
allowed  embeddings of the D-brane, with
no known origin in string theory.

A better idea is that it follows
from the usual quantization of $U(1)$ gauge field strength,
relevant because the D-brane is wrapped on $S^2$.
Indeed there is a well known mechanism of ``flux stabilization''
(which has been invoked in large extra dimension scenarios, for
example) which could then explain the brane's stability.
It is simply that the energy $\int F^2$ of a constant flux will be
inversely proportional to the volume, and thus the total energy with
the brane tension will have a minimum at non-vanishing volume.

This argument is not really correct, because, unlike what happens in the 
Maxwell energy, the Born-Infeld energy of a constant flux stays finite as 
the brane shrinks to zero volume (this is why there are no stable spherical 
D-branes in flat spacetime). What enters  in the D-brane energy, on the other 
hand, is the flux of  the  gauge-invariant combination ${\cal F}=
B+ 2\pi\alpha^\prime F$. In a varying external $B$ field the total energy 
including the brane tension can indeed have a minimum at nonzero volume, as 
we will show.

Our main result is to show that this explanation works not only
qualitatively but quantitatively: indeed, up to the well known
one loop shift $k\rightarrow k+2$ which 
 renormalizes the radius of $S^3$, computations
starting from the Born-Infeld action precisely reproduce the masses, 
multiplicities, and even the spectrum of small fluctuations of
these branes, as calculated in CFT. The  exact agreement
implies that  higher-order corrections to the Born-Infeld theory must
vanish -- this could be related to  the BPS property of the
corresponding  objects
in the fivebrane or $S^3\times AdS_3$ geometries.

 We consider the above results convincing evidence for the advocated
explanation of stability. They do in particular confirm the fact that
it is the $U(1)$ flux $\int F$  
(rather than  the flux of ${\cal F}$)  that must be quantized. 
This leads, however, to  an apparent paradox:
the RR charges of the   branes are not quantized.
We will explain  why the $F$-flux quantization is correct, and discuss
possible resolutions of the paradox in the final section.

\section{Semiclassical brane solutions}

Consider the WZW model on the group manifold of SU(2) -- we will comment on 
the generalization to other groups  later. A general group element can be
parametrized as 
$U = {\rm exp}(i\vec\psi \cdot \vec\sigma)$, where  $\vec \psi$ is a
 3-vector of
length $\psi$ pointing in the direction  ($\theta, \phi$), and
$\vec \sigma$ are the usual  Pauli matrices. The coordinate $\psi$
 takes values in the interval 
$[0,\pi]$ with the two extremes corresponding to the two elements of
 the center. 
In these coordinates the  metric and Neveu-Schwarz three-form  backgrounds
  read
\begin{equation}
ds^2 = k\alpha^\prime \left[ d\psi^2 + {\rm sin}^2\psi \Bigl( d\theta^2 +
{\rm sin}^2\theta\;  d\phi^2 \Bigr)\right]\ ,
\end{equation}
and
\begin{equation}
H  \equiv {dB} = {2k\alpha^\prime}\; {\rm sin}^2\psi\; 
{\rm sin}\theta\;
d\psi\; d\theta\; d\phi\ ,
\end{equation}
with  $k$  the (integer) level of the associated current algebra. 
We can choose a gauge  in which  the NS two-form is proportional to the
 volume  form of the
two-sphere spanned by $(\theta,\phi)$,  
\begin{equation}
B =  k\alpha^\prime 
\left( \psi - {{\rm sin}2\psi\over 2}\right)\; 
{\rm sin}\theta\; d\theta\; d\phi \ . 
\end{equation}
This is a smooth choice everywhere except at the point $\psi=\pi$.
The wavefunction of a fundamental string wrapping around this potential
singularity  picks up a Bohm-Aharonov phase equal
to $\int_{S^2} B/2\pi\alpha^\prime = 2\pi k$. The singularity is therefore
unobservable for integer $k$ as it should be.

   Let us next put this WZW model together with seven flat space-time
coordinates, so that the full geometry is $S^3\times R^7$. This is a
non-critical background for type-II string theory because the central
charges don't add up to ten, but the dilaton tadpole will not affect
our discussion of D-branes at leading order in the string-loop expansion.
 Consider now a static D2-brane
wrapping the $(\theta,\phi)$ two-sphere at
fixed value of  $\psi$. This configuration breaks the $SU(2)_L
\times SU(2)_R$ symmetry of the background to a diagonal $SU(2)$. 
If the dominant brane energy were tensive our configuration would
tend to shrink   to a point 
at one of the two poles of $S^3$, either  $\psi=0$ or $\psi=\pi$.
The total brane energy, on the other hand, has  contributions also from the
induced NS-NS  two-form $\hat B$ and from  the worldvolume
gauge field $F=dA$, which enter through  the  
invariant combination ${\cal F} = \hat B +2\pi\alpha^\prime F$. 
Consistently with the symmetry we may  
turn on   a uniform worldvolume  flux, 
\begin{equation}
F = dA = -{n\over 2}\; {\rm sin}\theta\; d\theta d\phi\ 
\end{equation}
where   $n$ is the `magnetic monopole' number.
For $0<n<k$ one can check that 
$\vert {\cal F} \vert$ is  locally maximum at the  poles, 
so this could  prevent  the D2-brane from collapsing. 

A crucial point in the further considerations is that it is the flux of $F$,
rather than that of $\cal F$, which is quantized.
This may seem counterintuitive in that $F$ is not invariant
under the gauge transformations $\delta \hat B=2\pi\alpha' d\Lambda$ and
$\delta A = -\Lambda$.
One might have expected 
the quantization condition to apply to the gauge invariant $\cal F$.

A first comment one can make is that gauge transformations for which
$\Lambda$ is single-valued do not affect $\int F$, so claiming 
that $\int F$ is quantized is not evident nonsense.
Although $\Lambda$ need not be single-valued, in fact such
large gauge transformations can only shift $\int F$ by an integer.
This is how the usual quantization condition on $\int H$ arises in
space-time language: one must define $\hat B$ in patches on $S^3$,
and the allowed transition functions between the patches are those
respecting this quantization condition.

This shows that the claim that $\int F$ and not $\int \cal F$
is quantized is sensible, but does not really prove it.
Indeed, any argument for this claim which starts
from a conventional world-volume gauge theory (such as the Born-Infeld
action) would be circular, as the conventional gauge potential
only makes sense if $\int F$ is quantized.

As is by now well-known, there do exist other gauge theories such as
noncommutative gauge theory in which  $\int F$ is not quantized in the usual way.
However, the examples in which this is known to make sense at present are 
related to manifolds with non-trivial fundamental group, such as the torus.
Indeed the case of $S^2$ has been much studied and the only noncommutative
gauge theories which are known to make sense in this case are based on the
``fuzzy sphere'' \cite{madore} and have finitely many degrees of freedom
(the original algebra of functions on $S^2$ is truncated).
There is even a ``no-go'' theorem \cite{Rieffel} to the
effect that deformations of the algebra of functions which do not make
this truncation, and which respect the natural $SO(3)$ symmetry, 
do not exist as bounded algebras.

Although this is a theorem, we have not proven that its assumptions are the 
physically appropriate ones, and so we are not claiming at this point
to show that noncommutative gauge theory
with a field theoretic number of degrees of freedom
does not exist on $S^2$.  The point of this discussion is to explain why
there is no sensible candidate low energy theory 
(at this writing) to describe the alternate hypothesis that $\int F$
is not quantized.

In any case, our hypothesis that $\int F$ is quantized
will be confirmed shortly
by the beautiful agreement of our results with those of
conformal field theory.

  Let us now fill in the appropriate formulae.   
The energy of our  D2-brane with
 $n$ units of worldvolume magnetic flux   in the
semiclassical  (large-$k$)  limit reads
\begin{eqnarray}
 E_n(\psi) & = & T_{(2)}\int_0^\pi  d\theta \int_0^{2\pi}
d\phi \; \sqrt{{\rm det}\; 
( \hat G + {\cal F}) } + \cdots \nonumber \\ 
&=& 4\pi k \alpha^\prime\;  T_{(2)}
\left( {\rm sin}^4\psi + (\psi -{{\rm sin}2\psi\over 2} -
 {\pi n\over k})^2\right)^{1/2}
 + \cdots   
\end{eqnarray}
where $T_{(2)}$ is the D-brane tension and 
the dots stand for  higher $\alpha^\prime$ corrections
(see for instance \cite{BBG}). For $0<n<k$ this expression has a unique
minimum away from the poles, at 
\begin{equation}
 \psi_{n} = {\pi n\over k}\ , 
\label{position}
\end{equation}
where it takes the value
\begin{equation}
M_n = 4\pi k\alpha^\prime \; T_{(2)} \; {\rm sin}{\pi n\over k}\ . 
\label{energy}
\end{equation}
For values of $n$ outside this range the minimum of the energy 
 is at $\psi=0$
(if $n<o$) or $\psi=\pi$ (if $n>k$) and it corresponds to a singular
configuration of the brane. 
This gives a total of $k-1$ non-singular configurations. In order to take into account
the well-known one loop shift of the sphere curvature, we should
replace everywhere $k$ by $k+2$. 
 There is however
 no  reason at this point to trust our expressions beyond the large
$k$ and $n$ (with $n/k$ held fixed) limit, since only in this limit
are the wordvolume
curvatures  small in string units.

 We can also  evaluate the
 D-particle charge induced by the background flux on the above
stable non-degenerate D2-branes. Using the standard formulae
\cite{revs} one finds  
\begin{equation}
 Q_n = T_{(2)} \int_{S^2} {\cal F} \  = \  
2 \pi k\alpha^\prime \; T_{(2)}\; {\rm sin}{2 \pi n\over k}\ . 
\label{charge}
\end{equation}
These charges are not even 
rationally related to each other -- here  is the apparent paradox we
have alluded to in the introduction. Note also that 
 in the flat limit,  $k\to\infty$ with $n$ held fixed,
eqs. (\ref{energy}), (\ref{charge})
 reduce  to the mass and charge of $n$ free
D-particles, 
$E_n \simeq  Q_n \simeq  n T_{(0)}$. The above stable configurations should
have  in fact a dual description as bound states of $n$ D-particles
on the sphere, which  can be analyzed along the lines of \cite{Myers}.
{}From  the exact properties of these bound states we can place
restrictions on the non-abelian Born Infeld theory, similar to those
of \cite{HT}\cite{Bain},  
but this is outside the scope of the present work.

\section{Small fluctuations}

    The `mini-superspace' analysis of  the previous section 
took only into account the degree of freedom corresponding to rigid
motions of the D2-brane in the $\psi$-direction. In this section we
derive the complete  spectrum of small quadratic fluctuations around the
above D-brane solutions. 
This will allow us to confirm their stability, to find their classical 
 moduli space,  and to compare later on 
with the   spectrum of boundary operators for  the corresponding 
Cardy states. 

We  use  static gauge in which the worldvolume is parametrised
by  $(t,\theta,\phi)$, and impose  $A_0=0$  for the worldvolume
gauge field. We ignore 
 for simplicity brane fluctuations and gauge-field components
 in the extra spectator 
spatial  directions, and concentrate on the three remaining
degrees of freedom 
\begin{equation}
\psi = {\pi n\over k} + \delta\  , 
\ \ \ A_\theta = {k\over 2\pi} \alpha_\theta\ , \ \ \ {\rm and}\ \ \ 
A_\phi = {n\over 2}({\rm cos}\theta -1) +
 {k\over 2\pi} \alpha_\phi \ . 
\end{equation}
Here the small fluctuations $\delta, \alpha_\theta, 
\alpha_\phi$ are arbitrary functions on $R\times S^2$, and the $k/2\pi$
normalization is introduced for convenience.  
The Born-Infeld energy-density  reads
\begin{equation}
{\cal L}_{BI} = T_{(2)} \sqrt{-{\rm det}\;(\hat G +{\cal F})}
\end{equation}
where 
\begin{equation}
\hat G +{\cal F} = k{\alpha}^\prime  \left( 
\begin{array}{lllll}
-{ 1\over k{\alpha}^\prime } + (\partial_t \delta)^2 
& \quad &
\partial_t \delta \partial_\theta \delta +\partial_t\alpha_\theta
& \quad &
\ \partial_t \delta \partial_\phi \delta +\partial_t\alpha_\phi 
\cr
\partial_t \delta \partial_\theta \delta - \partial_t\alpha_\theta
& \quad &
\ {\rm sin}^2\psi+ (\partial_\theta\delta)^2
 & \quad &
\ \ \ \partial_\theta\delta\partial_\phi\delta 
+{\cal F}_{\theta\phi} 
\cr
\partial_t \delta \partial_\phi \delta -\partial_t\alpha_\phi 
 & \quad &
\ \ \partial_\theta\delta\partial_\phi\delta 
- {\cal F}_{\theta\phi} & &
 {\rm sin}^2\psi\sin^2\theta+ (\partial_\phi\delta)^2 \cr
\end{array} 
\right)
\end{equation}
with
\begin{equation}
{\cal F}_{\theta\phi} \equiv
 \left(\delta - {\sin (2\psi) \over 2}\right)\sin\theta +
 \partial_\theta\alpha_\phi - \partial_\phi\alpha_\theta \ .
\end{equation}
In expanding out the determinant
to quadratic order,  terms involving $\delta$ 
off the diagonal drop out. After some tedious but
straightforward algebra the Born-Infeld
lagrangian up to quadratic order takes the form
\begin{equation}
\begin{array}l
{\cal L}_{BI} \propto \ \ \ 
 2f {\rm cot}({n\pi\over k}) + k\alpha^\prime
 {\rm sin}\theta  \Biggl[
(\partial_t\delta)^2 + (\partial_t\alpha_\theta)^2 +
(\partial_t\alpha_\phi)^2/ {\rm sin}^2\theta \Biggr]
\\
\\
- (1/ {\rm sin}\theta) \Biggl[
(\partial_\phi\delta)^2+ 
\sin^2\theta (\partial_\theta\delta)^2+ 2\sin^2\theta \delta^2 
+ 4\sin\theta \delta f +f^2 
\Biggr] \ . 
\end{array} 
\label{action} 
\end{equation}
We have here denoted $f\equiv
\partial_\theta\alpha_\phi - \partial_\phi\alpha_\theta$ 
for short, and have dropped the leading, fluctuation-independent term
of the lagrangian as well as an irrelevant multiplicative constant.

 The  above expression starts out with a linear term, which seems to
contradict our assertion that we are expanding  around a classical solution.
This is however not the case. 
The linear term is proportional to the fluctuation of the integrated flux,
which must be set to zero because of  the quantization condition.  
This demonstrates explicitly that it is the magnetic flux that stabilizes
the D-brane. The quadratic terms in (\ref{action}) are furthermore
independent of the background solution we expand around. This means that
the spectrum of fluctuations is independent of $n$, in agreement with
the conformal field-theory result as we will see soon.

{}From the above lagrangian we derive the following linearized equations
for the fluctuation fields,
\begin{equation}
{d^2\over dt^2} \left( 
\begin{array}l
\delta \\ \alpha_{\theta} \\ \alpha_\phi
\end{array}
\right)\ = \ {\cal O} \left( 
\begin{array}l
\delta \\ \alpha_{\theta} \\ \alpha_\phi
\end{array} \right)\ , 
\end{equation}
where the operator-valued matrix is
\begin{equation}
{\cal O} = -{1\over k\alpha^\prime}\; \left(
\begin{array}{lllll}
\sqr77 + 2 &\ \ \ \  \quad & -{2\over {\rm sin}\theta} \partial_\phi & \quad & 
\ \ \ \ {2\over {\rm sin}\theta} \partial_\theta \cr
& \quad & & \quad &\cr
\ \ {2\over {\rm sin}\theta} \partial_\phi
& \quad & 
\ \ -{1\over {\rm sin}^2\theta} \partial_\phi^2
& \quad & 
\ \ {1\over {\rm sin}^2\theta} \partial_\phi\partial_\theta
\cr
& \quad && \quad &\cr
-{2\over {\rm sin}\theta} \partial_\theta
& \quad &
{\rm  sin}\theta\; 
\partial_\theta {1\over {\rm sin}\theta} \partial_\phi 
 & \quad & 
- {\rm  sin} \theta\; \partial_\theta
 {1\over {\rm sin}\theta} \partial_\theta\cr
\end{array}
\right)
\end{equation}
and 
\begin{equation}
\sqr77 = -{1\over {\rm sin}^2\theta} \partial_\phi^2 - 
{1\over {\rm sin}\theta}\;\partial_\theta\; {\rm  sin} \theta\;
\partial_\theta
\end{equation}
is the covariant Laplacian on $S^2$. The operator $\cal O$ has zero
eigenvalues corresponding to the unphysical longitudinal polarization
of the photon. We can  extract the  transverse polarization 
by combining the last two equations so as to express everything in terms
of the physical fluctuation $f$. After some algebra the answer is
\begin{equation}
{d^2\over dt^2} 
\left( 
\begin{array}l
\ \ \  \delta \cr 
 f/{\rm sin}\theta \cr
\end{array}
\right)\ = \ -{1\over k\alpha^\prime}\;
\left(
\begin{array}{lll}
\sqr66 +2 &\quad & 2\cr
\ \ 2\; \sqr66 &\quad & \sqr66\cr
\end{array}
\right)
 \left( 
\begin{array}l
\ \ \ \delta \cr  f/{\rm sin}\theta \cr 
\end{array} 
\right)\ .  
\end{equation}

  We can now readily diagonalize this operator by going to a basis of
spherical harmonics, 
\begin{equation}
\delta = \sum_{l=0}^\infty \sum_{m=-l}^l\; \delta_{lm}
Y_{lm}(\theta,\phi)
\ \ \ {\rm and} \ \ \ 
f = \sin\theta \; \sum_{l=1}^\infty \sum_{m=-l}^l\; f_{lm}
Y_{lm}(\theta,\phi)
\end{equation}
with the reality conditions $\delta_{lm}= \delta_{l\ -m}^*$ and
similarly for $f$. Notice that the  absence of the s-wave in the expansion
of $f$ guarantees  the flux quantization condition, as can be checked
using the orthonormality of the spherical harmonics.   
For $l=0$ there is therefore only the $\delta$-fluctuation, and its
frequency squared is $2/k\alpha^\prime$. For all other $l$ we need
to consider the matrix
\begin{equation} 
{1\over k\alpha^\prime}\;
\left(
\begin{array}{lll}
l(l+1) +2 &\quad & 2\cr
\ \ 2l(l+1) &\quad & l(l+1)\cr
\end{array}
\right)
\end{equation}
whose eigenvalues are $(l+1)(l+2)/k\alpha^\prime$ and 
$l(l-1)/k\alpha^\prime$. Putting it all together we thus have the
following spectrum of quadratic fluctuations
\begin{equation}
m^2 = j(j+1)/k\alpha^\prime \ , \ \ {\rm in \ reps.}\ \ 
(j-1)\oplus (j+1)\ , \ \ {\rm for}\  j=0,1,2,...
\end{equation}
with the understanding that only the spin-one  representation appears in
the special case $j=0$. This corresponds precisely to a triplet
of zero modes, corresponding to arbitrary rotations of the worldvolume
two-sphere inside $S^3$. All other excitations have positive mass,
confirming  the stability of our solutions.

\section{CFT analysis}

  Let us  now compare the results of the last two sections with
those of  conformal field theory. 
 In the diagonal-invariant bosonic
theory there  are  ($k+1$)  Cardy \cite{C} boundary  states  
 preserving  a SU(2) symmetry, 
\begin{equation}
\vert q  \gg _{C} \  \equiv \  
 \sum_{p=0}^{k} 
   {S_{q p} \over \sqrt{S_{0p}}  }\;   
\vert p\gg _{I}
\label{sum}
\end{equation}
where $\vert p\gg _I$ is the Ishibashi (character) state corresponding
to the chiral primary of spin $p/2$, and  
\begin{equation}
S_{ij} = \sqrt{2\over k+2}\;
{\rm sin}  \left(  { (i+ 1)(j+ 1)\pi \over k+2} \right)  
\end{equation}
is the modular-transformation matrix.

We want to identify these states with the semiclassical configurations
of section 2. We have seen that  there are $k-1$ stable non-degenerate D2-branes,
when the  $S^3$ radius is $\sqrt{k{\alpha}^\prime}$, and adding the
two D-particles at the north and south poles gives the correct total number.
Alternatively, if we take into account the quantum shift $k\to k+2$,
we find exactly $k+1$ non-degenerate D2 branes. These two ways of
counting are of course indistinguishable in the semiclassical large
$k$ limit, since it is hard to tell the difference between a
point-like D-particle and one with radius $\sim\sqrt{{\alpha}^\prime}$
(this is the radius of the worldvolume for $n=1$ units of flux). 
Adopting the latter  point of view makes, however, the Born-Infeld
results exact -- this may be  related to supersymmetry in the
fivebrane context, but we did not have any  reason to expect it a priori. 
To exhibit  this precise agreement of the formulae  we will 
 assume that
the $S^3$ radius is $\sqrt{(k+2){\alpha}^\prime}$, and  identify
\begin{equation}
\vert n-1\gg _C \ \ \leftrightarrow\ \ ({\rm flux}-n\ \ {\rm D2-brane})
\end{equation}
where  the flux $n$ takes  the values $1,2,..,k+1$.

 The mass of a Cardy state   
can be read off from
the coefficient of its  $p=0$ Ishibashi
component which has non-vanishing overlap with 
 the (seven-dimensional) graviton and dilaton,  
see for instance  \cite{HKMS}. To be more precise, the interaction energy
between two D-branes a distance $r$ apart
due to the exchange of the (seven-dimensional) graviton and dilaton
can be calculated along the lines of \cite{revs}
with the result\footnote{We can formally ignore the dilaton tadpole in this
calculation.} 
\begin{equation}
{\cal E}(r) =  2\kappa_{(7)}^2\; M_n^2\; \Delta_{(6)}(r)
\end{equation}
where $\kappa_{(7)}$ is the gravitational coupling in seven dimensions,
$M_n$ the mass of the D-branes and $\Delta_{(6)}(r)$ the six-dimensional
Euclidean Green's function. 
The string-theory calculation for this
interaction energy on the other hand is
\begin{equation}
{\cal E}(r) = (2\pi)^4 {\vert S_{n0}\vert^2\over S_{00}} \; \Delta_{(6)}(r)
\end{equation}
where we have here projected onto the identity-operator in the closed-string
channel of the amplitude, and we have set $\alpha^\prime = 1/2$.
Comparing the two expressions gives
\begin{equation}
M_n = {2\sqrt{2}\pi^2\over \kappa_{(7)}}\; {\vert S_{n0}\vert\over
\sqrt{ S_{00}}}\;  =\;  {(2\pi)^{3/2} (2k+4)^{1/4}
\over \kappa_{(7)}} \;  {\rm sin}\left({n\pi\over k+2}\right) \ ,
\end{equation}
where we have here used the expressions for  the S-matrix coefficients.
Using now the relations
\begin{equation}
T_{(2)} = {\sqrt{2\pi^3} \over \kappa_{(10)}}\ , 
\ \ {\rm and}\ \ \  \kappa_{(10)}^2 =
 \kappa_{(7)}^2  \left({k+2\over 2}\right)^{3/2} 2\pi^2
\end{equation}
one can verify that the above mass agrees precisely with our
semiclassical result (\ref{energy}),
 including all the numerical prefactors.

 Consider next  the spectrum of quadratic fluctuations, to   be compared
with the open-string excitations in the ${\cal H}_{(n-1)(n-1)}$ Hilbert space.
If we neglect transverse spatial dimensions, the light
states in this Hilbert space are of the form
\begin{equation}
 \vert {\rm open} > =  J^a_{-1} \vert j>
\label{states}
\end{equation}
where $J^a$ are the SU(2) currents and $\vert j>$ is created by
a primary field with  $j= 0,\cdots k/2$. These transform in the
$(j-1)\oplus j\oplus (j+1)$ representations of $SU(2)$, but imposing
the (super)Virasoro constraint will project the representation $j$ out
of the spectrum. One way to see this is to note that there are as many
constraints as number of primaries, namely $2j+1$, and since physical
states must form $SU(2)$ representations it is necessarily the $j$
representation that is projected out. 
The conformal weight of the  vertex operators corresponding to the
states (\ref{states}) is  
$j(j+1)/(k+2)$ , in complete agreement with the semiclassical mass
formula  of small fluctuations derived in section three.

  Another qualitative confirmation of the results  in section two follows from 
an analysis  of the `wavefunctions' of the Cardy states in position space \cite{FS}.
These are peaked around equally-spaced values of the polar angle 
$\psi$, in agreement again with the semiclassical  result (\ref{position}).   
Notice that the moduli space for 
rigid translations of the  D2-branes on the group manifold  corresponds to the
freedom of obtaining  equivalent Cardy states by 
group conjugation.

Finally, let us compare   the induced D-particle  charge (\ref{charge})
with the result of CFT. In the CFT this charge is
given by the $p=1$ coefficient of the  Cardy state, because
the corresponding closed-string  RR  states transform in
 the $(p/2\otimes 1/2,p/2\otimes 1/2) $ 
representation of $SU(2)_L\times SU(2)_R$. The reason is that the 
 zero modes
of the supersymmetric WZW fermions are realized on a bispinor of $SU(2)_L\times SU(2)_R$.
Now the D-particle charge of interest  is
a $SO(4)$ singlet -- this can be verified explicitly by checking that
it does not transform under  rigid translations  of the D2-brane on $S^3$.
Thus only $p=1$ contributes to this coupling,  and a calculation
similar to the one for the mass gives 
\begin{equation}\label{rrcharges}
Q_n =\;  {(2\pi)^{3/2} (2k+4)^{1/4}
\over 2 \kappa_{(7)}} \;  {\rm sin}\left({2n\pi\over k+2}\right) \ ,
\end{equation}
in perfect agreement again with the result of section two. 
That these charges are not rationally related to each other
is  thus confirmed by the CFT analysis -- we will return to the point
in the final section.  

\vfil\eject

\section{General group manifolds}

In this section we discuss some aspects of the generalization of our
results to compact Lie groups $G$ which we assume to be simple, connected and,
for simplicity, to be simply connected.
Our discussion will be entirely topological, the precise form of the
metric and antisymmetric tensor will not play any role. The reader not
interested in this generalization can go directly to the final section.

The D-brane world volumes for all boundary conditions 
for which the gluing of left movers and right movers at the boundary is 
given by an automorphism $\omega$ of $G$ have been described in \cite{FS}.
They are (regular) twined conjugacy classes, i.e.\ they are subspaces of the 
form
\begin{equation} 
{\cal C}_\omega(g) = \{ h g \omega(h)^{-1} \,\mbox{ with } h\in G \} 
\end{equation}
where $g\in G$ is a regular element.

Let us describe the geometry of twined conjugacy classes in somewhat more
detail: for any automorphism $\omega$, there is a maximal torus $T$ of $G$ 
that is invariant under $\omega$. The subgroup of elements of $T$ that
are left pointwise fixed by $\omega$, 
\begin{equation}
 T^\omega = \{ t\in T | \omega(t) =t \} \,,
\end{equation} 
is not a torus, but a semi-direct product of a torus and a finite
abelian group; its connected component $T^\omega_0$ of the identity is a 
torus. In case $\omega$ is an inner automorphism -- and this is always true
for $G= SU(2)$ -- all subgroups of $G$ coincide:
\begin{equation}
 T = T^\omega_0 = T^\omega \, .
 \end{equation} 
For inner automorphisms, this torus actually coincides with a maximal
torus and the dimension of the torus equals the rank of $G$; so e.g.\ for 
inner automorphisms of $SU(3)$ we have a two-dimensional torus. For outer 
automorphisms, the dimension of $T^\omega_0$ is smaller than the rank;
for outer automorphisms of $SU(3)$ it is equal to one.

Weyl's classical theory of conjugacy classes has in fact a nice 
generalisation to twined conjugacy classes. We will sketch some 
statements
of this theory. The central tool  is the following:
given a maximal torus $T$, we define a map from the coset space $G/T$ and
the maximal torus $T$ to the group by using conjugation:
\begin{equation}\begin{array}l
q\,: \quad G/T \times T \to G  \\
q( [g],t) = g t g^{-1} 
\end{array}\end{equation}
Weyl could show that the mapping degree of $q$ equals the number of elements
in the Weyl group $W$. Maps with positive degree are surjective, so
in particular one sees that any element of $G$ is conjugated to some element
in the maximal torus $T$. So any conjugacy class can be characterized by
an element of the maximal torus. In fact, different elements of the
maximal torus parametrize identical conjugacy classes if and only if they
are related by the action of the Weyl group. In the case of $SU(2)$, 
a maximal torus is one-dimensional; examples are given by circles of
constant values of $\phi$ and $\theta$; they can be parametrized by $\psi$. 
The Weyl group is just $Z_2$, and its action has been taken into account
by restricting   $\psi$ to the range $0\leq\psi\leq \pi$.
Finally, fixing $t$ we see that regular conjugacy classes are isomorphic to 
the homogeneous space $G/T$. In the case of $SU(2)$ this gives
$SU(2)/U(1)$ which is isomorphic to the two-sphere.

The results nicely generalize to twisted conjugacy classes (for details
see \cite{FS}). For any automorphism $\omega$, $q$ is replaced by
$q_\omega$ that is defined via twisted conjugation:
\begin{equation}\begin{array}l
q_\omega\,: \quad G/T^\omega_0 \times T^\omega_0 \to G  \\
q_\omega( [g],t) = g t \omega(g^{-1})
\end{array}\end{equation}
The mapping degree of $q_\omega$ can be shown to be positive. To state
more precise results, we need the subgroup $W_\omega$ of the Weyl group
$W$ that commutes with the action of $W$ on the weight space:
\begin{equation}
W_\omega := \{ w\in W | \omega^* w = w \omega^* \mbox{ for all }
w\in W\} 
\end{equation}
The group $W_\omega$ has been shown in \cite{FURS} to be isomorphic to the 
Weyl group of the so-called orbit Lie algebra \cite{FSS}. For the
outer automorphism of $SU(3)$ this group is $Z_2$, which is the Weyl group
of the orbit Lie algebra $SU(2)$.

The mapping degree of $q_\omega$ is just $n_{T^\omega} |W_\omega|$, where
$n_{T^\omega}$ is the number of connected components of $T^\omega$. 
Weyls classical results can now be generalized to twined conjugacy classes.
All statements remain true, provided one replaces the maximal torus $T$ by 
$T^\omega_0$ and the Weyl group $W$ by $W_\omega$: Twined conjugacy classes
are characterized by elements of $T^\omega_0$; different elements of
$T^\omega_0$ describe identical twined conjugacy classes if and only if
they are related by the action of $W_\omega$. Regular twined conjugacy classes 
are isomorphic to the homogeneous space $G/T^\omega_0$. Even a generalization
of Weyl's integration formula holds \cite{wendt}.

To give an explicit example, regular D-branes for inner automorphisms of
$SU(3)$ are isomorphic to $SU(3)/U(1)^2$ and are thus six-dimensional.
They are characterized by two parameters. For {\em outer} automorphisms,
they are isomorphic to $SU(3)/U(1)$ and therefore seven-dimensional.
For their characterization a single parameter suffices. Outer automorphisms
therefore change the dimensionality of the  worldvolume.

Extending the analysis of $SU(2)$ to  other groups 
 requires a detailed knowledge of the differential geometry of the
group manifold, in particular a good choice of coordinates. The corresponding
calculations become rather complicated and are beyond the scope of the
present note. 
However, there is a simple and yet non-trivial check 
of the stabilization mechanism: we expect  as many
independent $U(1)$ fluxes as the number of transverse brane coordinates that
must be stabilized.

The possible $U(1)$
fluxes  on the worldvolume of the D-brane
are  given by
\begin{equation}
 {\rm dim} H^2( G/T^\omega_0,{\rm R} )
 \end{equation} 
Following our previous discussion, the description of a specific D-brane 
requires on the other hand $\dim T^\omega_0 $ parameters.
We should thus expect  the general relation
\begin{equation} 
{\rm dim} H^2( G/T^\omega_0,{\bf R} )
  = {\rm dim}  T^\omega_0 
\label{quant} 
\end{equation}

Such a relation does indeed hold in full generality: for a simply connected 
compact Lie group $G$ also the second homotopy group $\pi_2(G)$ vanishes.
The long exact sequence in homotopy
\begin{equation}
 \ldots \to \pi_k(G) \to \pi_k(G/T^\omega_0) \to \pi_{k-1}(T) 
\to \pi_{k-1}(G) \to \ldots
\end{equation}
implies for $k=1$ that the homogeneous space $G/T^\omega_0$
is simply connected and for $k=2$ that $\pi_2(G/T^\omega_0)$ is
isomorphic to $\pi_1(T^\omega_0)$. The latter is a free abelian
group whose rank is $\dim T^\omega_0$. The homotopy group $\pi_2(G/T^\omega_0)$
therefore coincides with the homology we want to determine, and we
find indeed that
\begin{equation}
 H^2( G/T^\omega_0,R ) \cong  \pi_2(G/T^\omega_0)
\cong  \pi_1(T^\omega_0) \cong  {\bf Z}^{{\rm dim}  T^\omega_0} 
\end{equation}
Notice in particular that the line bundles over $G/T^\omega_0$
do not have any continuous parameters. This generalizes the situation
of $SU(2)$, where we consider bundles over $S^2$. This is reflected
in the conformal field theory analysis by the fact that we find
D-brane worldvolumes whose only continuous deformations are given by
(inner) automorphisms of the group, but which do not have any other moduli.

The fact that the relation (\ref{quant}) always holds shows
that  the advocated  mechanism could indeed be responsible for the stability
of all known WZW D-branes.

\section{Fivebrane and a paradox}

To further discuss the physics of the $SU(2)$  branes, let us consider
a configuration of $N$ coincident supersymmetric (NS) five-branes in
type II theory.

The full fivebrane background is (in string frame)
\begin{eqnarray}\label{fivebranebkgd}
ds^2 &=& dx^2 + f(r) dy^2 \cr
e^{2\phi} &=& g_s^2 f(r) \cr
f(r) &=& 1 + {N\alpha'\over r^2} \cr
H &=& N\alpha' \epsilon_3
\end{eqnarray}
where $x$ are the $5+1$ longitudinal coordinates,
$y$ are $4$ transverse coordinates and $r=|y|$.

In the near-horizon limit $r\rightarrow 0$,
the background factorizes into a radial component and an $S^3$.
The corresponding CFT has a Feigin-Fuchs (linear dilaton)
field in addition to the supersymmetric SU(2) WZW model \cite{CHS}. 
The supersymmetric WZW model can be realized as a tensor product of three
free fermions (with level $2$ current algebra) and a bosonic WZW
model of level $k$ (the $k$ of the previous sections).
The fivebrane number $N$ is identified with the total
central charge $k+2$ of the $SU(2)$ current algebra.
This may sound  unsatisfactory,  as there
is apparently no candidate theory for a single five-brane,
 but it  agrees  with the standard lore 
that the center-of-mass degrees of freedom of the branes 
are not be visible in the dual holographic theory. A single
fivebrane has no degrees of freedom other than center-of-mass, and
hence no dual holographic description.

The D particles of the previous section are nonsupersymmetric and
unstable in this background. They are momentum modes in the eleventh
dimension, which tend to fall towards the core of the fivebrane where
the eleventh dimension blows up and the D-particles become massless.
This agrees with the expectation that they should complete SO(5)
representations of the fivebrane fields \cite{ABKS}. 

D$2$-branes which end on the fivebrane are supersymmetric and correspond
to the product of a Neumann boundary state in the linear dilaton theory,
and a Dirichlet (D$0$) boundary state in the WZW model.
In this context, the additional WZW boundary states will correspond to
D$4$ branes extending along the radial direction (and so ending on the
fivebranes), but now (in the near horizon regime) ``wrapped'' on an
transverse $S^2$, forming a conical geometry.  

To study the supersymmetry properties of these branes, we need to
write down the space-time supersymmetry generators.
The world-volume supersymmetry generators are given in \cite{CHS} and
the D$2$ boundary conditions in the WZW model preserve an $N=4$
world-sheet supersymmetry.  This is enough to guarantee that the
world-volume operator corresponding to the space-time supercharge
also exists with these boundary conditions.

The conjugate brane (in the sense of electric-magnetic duality) would be
a D$4$-brane with three dimensions wrapped on $S^3$ and $1+1$
extending in Minkowski space.  Although this of course does not exist
because it is wrapping a surface with $H\ne 0$, there is a
similar object which is believed to exist \cite{ABKS}.
The total flux $\int H$
on the brane can be made zero by allowing $k$ D$2$-branes to end on the
brane (and extend outward), analogous to the ``baryon'' of \cite{witten}.

The existence of the conjugate object would appear to
require D$2$ charge quantization. So why is RR charge not quantized ?  

We first note that there is a
superficially similar effect, already visible in toroidal compactification,
with a much simpler explanation.
Since the integrals $\int \hat B$ over two-cycles in the target space
can take arbitrary non-zero values, the induced RR charges 
$\int C\wedge \hat B$ are non-integral.  
However, this is not a violation
of quantization but rather a rotation of the entire charge lattice.

A simple way to see that this is not what is happening here is to realize
that charge quantization requires that there be
an integral basis of the charge lattice of rank equal to the number
of charges (one must then check that the DSZ form is integral of course).
In the present case (and for $SU(2)$), 
we have found  $O(k/2)$ distinct `charges'  satisfying no
integer relations, but at most  two independent RR charges 
(from the two-dimensional cohomology of $S^3$).

  The way one avoids an immediate  contradiction is by noting that the RR
fields  in question are  massive in the near-horizon geometry.   
This can be seen in the CFT where the normalizable vertex operators 
have (six-dimensional) mass bounded below by the background charge of
the Feigin-Fuchs coordinate \cite{ABKS}. Alternatively, in the 
low energy effective theory, this follows from the 
 Chern-Simons coupling $\int G\wedge G\wedge \hat B$.
We work in the usual string conventions
in which the RR kinetic terms and sources are independent of the dilaton.
This leads to
\begin{equation}\label{rreqn}
d*G = H \wedge G + \delta^{(7)} + B \wedge \delta^{(5)}
\end{equation}
and
\begin{equation}\label{rrdualeqn}
dG = \delta^{(5)}
\end{equation}
where $G=dC^{(3)}$ is the four-form field strength, $\delta^{(9-p)}$
is the source associated to a $p+1$-brane (a $9-p$ form normal to the
world-volume).
The conserved electric and magnetic RR charges are then
$$Q_M = \int G$$
and
$$Q_E = \int *G - B\wedge G.$$
In the near-horizon limit of the five-brane background, 
$B$ is independent of distance $r$.  In this case
the CS term makes the RR field effectively massive, 
so the quantization condition will not be visible.

In the true five-brane background, when we go to asymptotic infinity
(the Minkowski region),
the $B$ field does fall off with distance, and the charge quantization
must become observable again.  This regime is of course not described
by our explicit CFT.  Since the volume of the $S^3$ grows with radius
in the normal way here, presumably the conical four-brane must asymptote
to a cylindrical four-brane (or even $n$ two-branes again).
This fits with the asymptotic
BPS bound which implies that the tension here
is just $n$ times that of the original two-brane.

In a general background with non-zero $H$,
it is clear that the induced RR charge will depend on the embedding
of the brane through $\int \hat B$ and so this contribution cannot
be quantized.  However this variation could be  cancelled by the bulk CS term:
the total variation of the right hand side of
(\ref{rreqn}) under a variation of the embedding of the D$4$-brane
is zero, under a suitable interpretation of the boundary terms.

These  points seem to us to resolve the paradox both in the context of
the near-horizon geometry (because the RR fields are massive) and in the
full geometry (in which the $H$ flux falls off fast enough to apply the second
argument).
However they leave open the interesting question of just what
the CFT results  (\ref{rrcharges}) are  measuring. One possibility is that
they are related to the $N-1 = k+1$ independent charges of the 
holographically dual field theory on the fivebranes. For $N$ type IIB
fivebranes the worldvolume  theory has $SU(N)$ gauge symmetry,
 and hence $N-1$
independent charges. These charges may indeed
correspond  to the different allowed couplings
of the  massive RR field in the bulk theory.

We feel there is more to say
about this issue, but will leave it for future work.

\vskip 2cm
{\bf Acknowledgements}

The authors thank the organizers of the workshop on
`Non-Commutative Gauge Theory' at  the Lorentz Center in  Leiden, Holland, 
where  this collaboration started. CB thanks the New  Center for High Energy 
Physics at the University of Rutgers for hospitality during completion of this 
work. MRD would like to thank A.\ Rajaraman and M.\ Rozali for discussions on
this problem and especially for emphasizing the importance of
RR charge quantization. We also thank O.\ Aharony, J.\ Maldacena and 
J.\ Polchinski for useful comments.

%%%%%%%%%%%%%%%%%%%%%%%%%%%%%%%%%%%%%%%%%%%%%%%%%%%%%%%%

\end{document}